\documentclass[11pt]{article}
\usepackage[margin=1in]{geometry}
\usepackage[colorlinks=true, pdfborder={0 0 0}, citecolor=blue, linkcolor=blue]{hyperref}
\usepackage{graphicx}
\graphicspath{ {./Figures/} }
\usepackage{authblk}
\renewcommand\Authfont{\normalsize}
\renewcommand\Affilfont{\itshape\footnotesize}
\usepackage{siunitx}
\usepackage{chemformula}
\usepackage{url}
\usepackage[labelfont=bf]{caption}
\usepackage{setspace}
\usepackage{upgreek}
\usepackage{helvet}
\usepackage{sfmath}
\usepackage[numbers,square,sort&compress]{natbib}
\usepackage[resetlabels]{multibib}
\newcites{Sup}{References}
\usepackage[acronym]{glossaries}
\glsdisablehyper

\title{{\Large \textbf{Scalable graphene platform for Tbits/s data transmission}}}

\author[1]{Brian S. Lee}
\author[2]{Alexandre P. Freitas}
\author[1]{Andres Gil-Molina}
\author[1]{Euijae Shim}
\author[3]{Yibo Zhu}
\author[3]{James Hone}
\author[1,*]{Michal Lipson}
\affil[1]{Department of Electrical Engineering, Columbia University, New York, NY 10027, USA}
\affil[2]{School of Electrical and Computer Engineering, University of Campinas, Campinas-SP, 13083-970, Brazil}
\affil[3]{Department of Mechanical Engineering, Columbia University, New York, NY 10027, USA}
\affil[*]{Corresponding author: ml3745@columbia.edu}
\date{}

\newacronym{sin}{\ch{Si3N4}}{silicon nitride}
\newacronym{soi}{SOI}{silicon-on-insulator}
\newacronym{al2o3}{\ch{Al2O3}}{alumina}
\newacronym{hafnia}{\ch{HfO2}}{hafnia}
\newacronym{sio2}{\ch{SiO2}}{silicon oxide}
\newacronym{cmp}{CMP}{chemical mechanical planarization}
\newacronym{pecvd}{PECVD}{plasma-enhanced chemical vapor deposition}
\newacronym{lpcvd}{LPCVD}{low pressure chemical vapor deposition}
\newacronym{ald}{ALD}{atomic layer deposition}
\newacronym{ber}{BER}{bit error rate}
\newacronym{hdfec}{HD-FEC}{hard-decision forward error correction}
\newacronym{nrz}{NRZ}{non-return-to-zero}
\newacronym{prbs}{PRBS}{pseudorandom binary sequence}
\newacronym{gfet}{GFET}{graphene field-effect transistor}
\newacronym{cvd}{CVD}{chemical vapor deposition}
\newacronym{osnr}{OSNR}{optical signal-to-noise ratio}
\newacronym{eet}{EET}{elementary effect test}
\newacronym{ee}{EE}{elementary effect}
\newacronym{fde}{FDE}{finite difference eigenmode}

\newcommand{\vthreedb}{$\Delta$V$ _{3dB} $}
\newcommand{\um}{\textmu m}
\newcommand{\doping}{$ n_0 $}
\newcommand{\deldoping}{$ \Delta n_0 $}
\newcommand{\extdoping}{$ \Delta n_{ext} $}
\newcommand{\gbps}{Gbits/s}
\newcommand{\vpp}{V$ _{pp} $}
\newcommand{\kap}{$\kappa$}
\newcommand{\ohmum}{$ \Upomega \cdot$\textmu m}
\newcommand{\cmvs}{cm$^\text{2}$/V$\cdot$s}

\begin{document}
    \maketitle
    \vspace{-3em}
    \doublespacing

    \section*{Abstract}
   	\textbf{To date, no electro-optic platform enables devices with high bandwidth, small footprint, and low power consumption, while also enabling mass production. Here we demonstrate high-yield fabrication of high-speed graphene electro-absorption modulators using CVD-grown graphene. We minimize variation in device performance from graphene inhomogeneity over large area by engineering graphene-mode overlap and device capacitance to ensure high extinction ratio. We fabricate an 8 mm $ \times $ 1 mm chip with 32 graphene electro-absorption modulators and measure 94\% yield with bit error rate below the hard-decision forward error correction limit at 7 \gbps, amounting to a total aggregated data rate of 210 \gbps. Monte Carlo simulations show that data rates $ > $ 0.6 Tbits/s are within reach by further optimizing device cross-section, paving the way for graphene-based ultra-high data rate applications.}

    \section{Introduction}
    With data traffic growing exponentially, there is an urgent demand for optical modulators with large bandwidth, small footprint, and low power consumption, that can be mass-produced \cite{romagnoli2018graphene}. Although integrated graphene modulators offer high bandwidth with small footprint and low energy consumption \cite{li2013modulation, phare2015graphene, giambra2019high}, low device yield has been a major challenge. There have been significant advancements in growing and transferring \gls{cvd} graphene at wafer-scale \cite{li2009large, bae2010roll}, but these films are inherently polycrystalline and exhibit nonuniformities of conductance and carrier density at the micrometer scale due to grain boundaries and defects \cite{buron2012graphene, gong2012layer}. Furthermore, graphene transfer and fabrication steps contribute to nonuniformity by increasing contamination (such as polymer residues) and unwanted doping. While these inhomogeneities are believed to cause large variation in device performance \cite{rizzi2012cascading, heo2013graphene, theofanopoulos2019high}, large area analysis for photonic devices have not been shown to date.

\begin{figure}[!b]
    	\centering
    	\includegraphics[width=120mm]{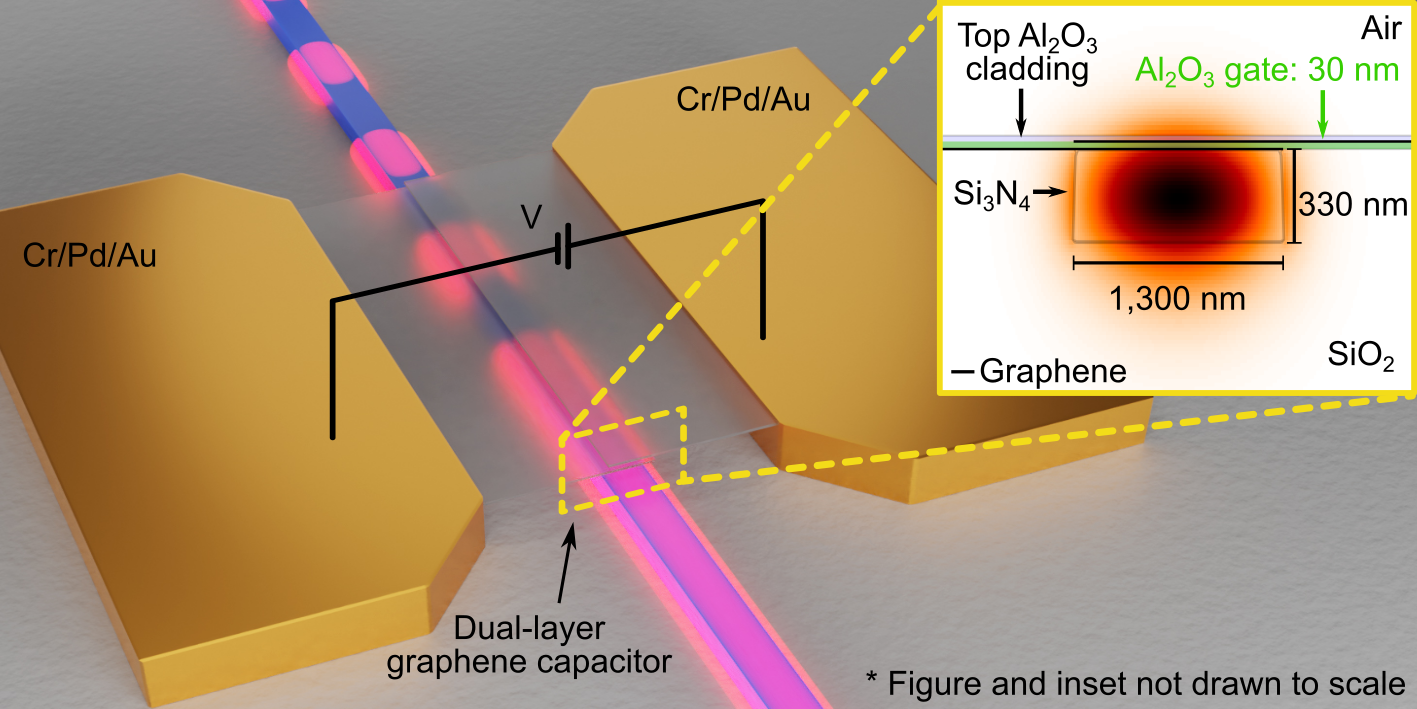}
    	\caption{Schematic of the graphene electro-absorption modulator.\\
    		Light guided by the waveguide passes through the modulator and is absorbed by the graphene capacitor. Inset: device cross-section consisting of two graphene sheets (black lines) separated by a 30 nm \gls{ald} \gls{al2o3} gate dielectric (green layer) and a 330 nm tall \gls{sin} waveguide. The widths of the waveguide and graphene capacitor are 1.3 \um. We superimpose the simulated fundamental quasi-TE mode of the waveguide with the device structures to show graphene capacitor's proximity to the mode for strong coupling via evanescent waves. We deposit a 30 nm \gls{al2o3} cladding above the top graphene sheet to protect the capacitor.}
    	\label{fig:schematics}
    \end{figure}

    Here we demonstrate large yield of high-speed graphene electro-absorption modulators using commercially available \gls{cvd} polycrystalline graphene. We design the modulator to exhibit high extinction ratio relative to the undesired doping induced extinction, ensuring high optical signal-to-noise ratio (OSNR). We achieve strong extinction ratio modulation by integrating a dual-layer graphene capacitor on a \gls{sin} waveguide with \gls{al2o3} gate dielectric for large device capacitance. Moreover, we ensure strong graphene-mode overlap by placing this graphene capacitor directly above the \gls{sin} waveguide for maximum graphene-mode coupling. We show the schematic of the graphene electro-absorption modulator in \autoref{fig:schematics}. Light guided by the waveguide is absorbed by the graphene capacitor.
	The capacitor consists of two graphene sheets (black lines in the inset of \autoref{fig:schematics}) separated by a 30 nm \gls{ald} \gls{al2o3} gate dielectric (green layer) to form a parallel plate capacitor. We modulate graphene absorption and propagation loss by applying voltage to the capacitor. This electrostatically gates the two graphene sheets which induces Pauli-blocking and suppresses interband transitions of carriers \cite{wang2008gate, liu2011graphene}. As shown in the inset of \autoref{fig:schematics}, we maximize the coupling between graphene and the fundamental quasi-TE mode via evanescent waves by placing the capacitor directly above the waveguide.

    \section{Graphene transmitter chip fabrication}
\begin{figure}[!b]
    	\centering
    	\noindent\makebox[\textwidth]{\includegraphics[width=183mm]{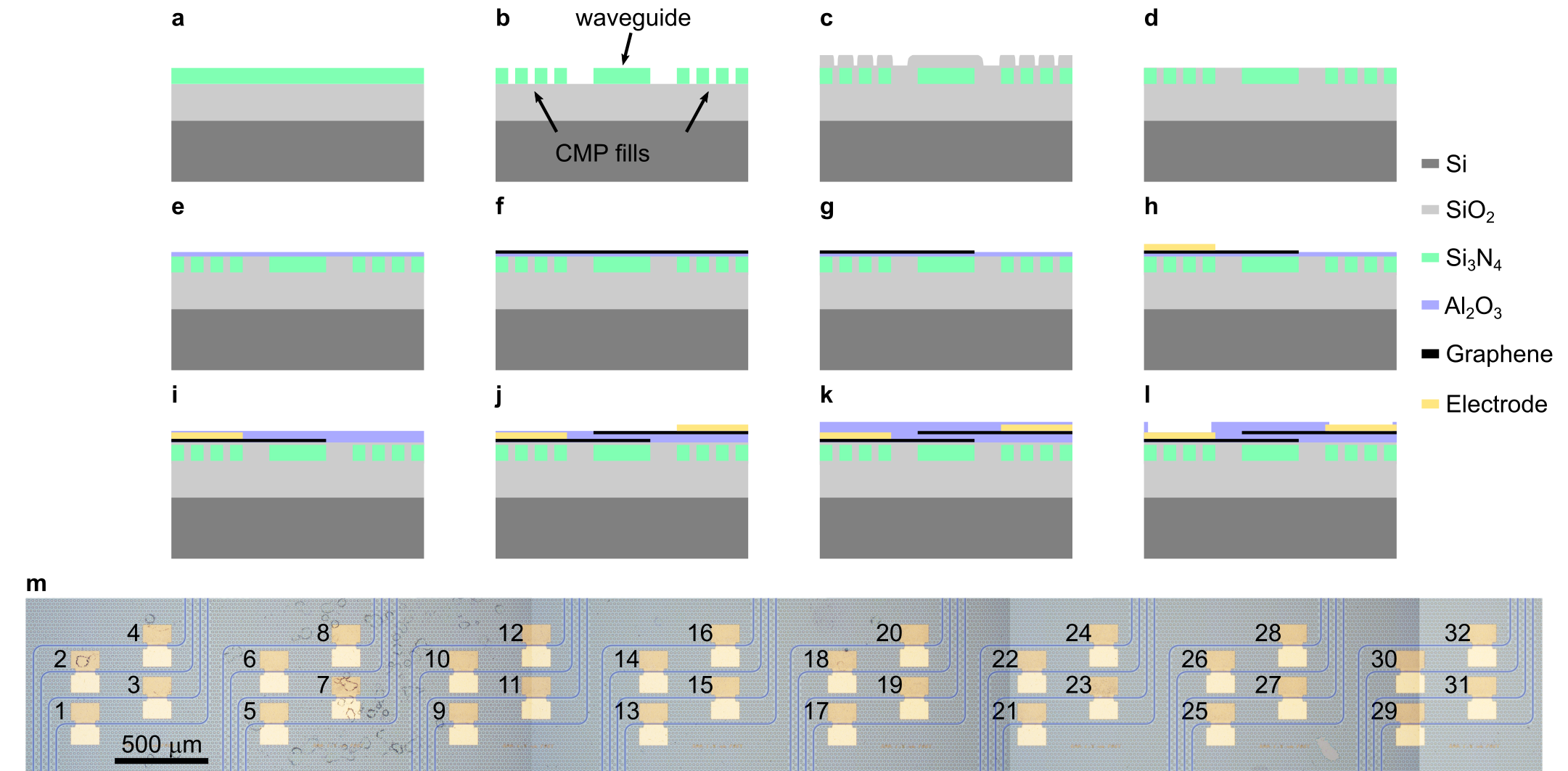}}
    	\caption{Fabrication of the graphene transmitter chip.\\
    		(a) Deposition of 330 nm of \gls{lpcvd} \gls{sin} on 4.3 \um-thick thermal \gls{sio2} on silicon substrate. (b) Patterning and etching of the waveguides and 10 \um\ by 10 \um\ square fills for \gls{cmp}. (c) Cladding deposition of \gls{pecvd} \gls{sio2}. (d) Top cladding removal and planarization of the chip surface with \gls{cmp} to expose the top of the \gls{sin} waveguide. (e) Deposition of 10 nm of \gls{ald} \gls{al2o3}. (f) Transferring of \gls{cvd} graphene on copper foil (Grolltex Inc.) via electrochemical delamination wet transfer (also see Supplementary Information). (g) Etching of the transferred graphene film with \ch{O2} plasma to form the bottom capacitor plate. (h) Deposition of the electrode (Cr/Pd/Au, 1 nm/45 nm/50 nm) using e-beam evaporation at high vacuum ($ 10^{-8} $ Torr). (i) Thermal evaporation of 1 nm of aluminum to serve as a seed layer (after electrode lift-off) for subsequent deposition of 30 nm \gls{ald} \gls{al2o3} gate dielectric. (j) Transferring and patterning of the second graphene layer similarly to the first one, followed by electrode deposition (steps f-h). (k) Deposition of a final of 30 nm thick \gls{ald} \gls{al2o3} on top of the second graphene sheet to protect the capacitor. (l) Opening vias with buffered oxide etch (50:1) to remove \gls{al2o3} and access the electrodes. (m) Tiled optical micrograph of the 8 mm $ \times $ 1 mm transmitter chip with 32 graphene electro-absorption modulators. The waveguides are false-colored in blue to stand out among CMP fills. Each device consists of a 100 \um\ long and 1.3 \um\ wide dual-layer graphene capacitor to modulate waveguide transmission. Therefore, the device's active area only covers approximately 370 \um$ ^2 $.}
    	\label{fig:fabrication}
    \end{figure}

    We fabricate an 8 mm $ \times $ 1 mm chip with 32 graphene electro-absorption modulators to characterize variations in modulation depth, bandwidth, and data transmission quality. We outline the fabrication steps for the graphene modulator in \autoref{fig:fabrication}.
	We first deposit 330 nm of \gls{lpcvd} \gls{sin} on 4.3 \um-thick thermal \gls{sio2} on silicon substrate. We then pattern and etch the waveguides and 10 \um\ by 10 \um\ square fills for \gls{cmp} (see \autoref{fig:fabrication}b). We deposit a cladding layer over the \gls{sin} patterns with \gls{pecvd} \gls{sio2}. We remove the top cladding and planarize the chip with \gls{cmp} to expose the top surface of the \gls{sin} waveguide. This planarization step helps prevent damage to the graphene sheets during transfer and ensures that the graphene is in direct contact with the waveguide, maximizing mode overlap. To screen charge impurities on the surface and minimize undesired substrate doping to graphene, we deposit 10 nm of \gls{al2o3} via \gls{ald}. We transfer \gls{cvd} graphene on copper foil (Grolltex Inc. \cite{grolltex}) via electrochemical delamination wet transfer. This method enables transfer of large-area films with ease and minimal chemical/mechanical damage to the graphene \cite{verguts2018graphene, zhu2018monolayer} (see Supplementary Information). We etch the transferred graphene film with \ch{O2} plasma to form the bottom capacitor plate, followed by deposition of the electrode (Cr/Pd/Au, 1 nm/45 nm/50 nm) using e-beam evaporation at high vacuum ($ 10^{-8} $ Torr). After electrode lift-off, we thermally evaporate 1 nm of aluminum to serve as a seed layer for subsequent deposition of 30 nm \gls{ald} \gls{al2o3} gate dielectric. We transfer and pattern the second graphene layer similarly to the first one, followed by electrode deposition (\autoref{fig:fabrication}f-h). In order to protect the capacitor, we deposit a final layer of 30 nm thick \gls{ald} \gls{al2o3} on top of the second graphene sheet. We open vias using buffered oxide etch (50:1) to remove \gls{al2o3} and access the electrodes. The fabricated graphene transmitter chip is shown in the tiled optical micrograph in \autoref{fig:fabrication}m. The waveguides are false-colored in blue to stand out among CMP fills. Each device consists of a 100 \um\ long and 1.3 \um\ wide dual-layer graphene capacitor used to modulate waveguide transmission.

    \section{Experimental Results}
    \begin{figure}[!t]
        \centering
        \includegraphics[width=163mm]{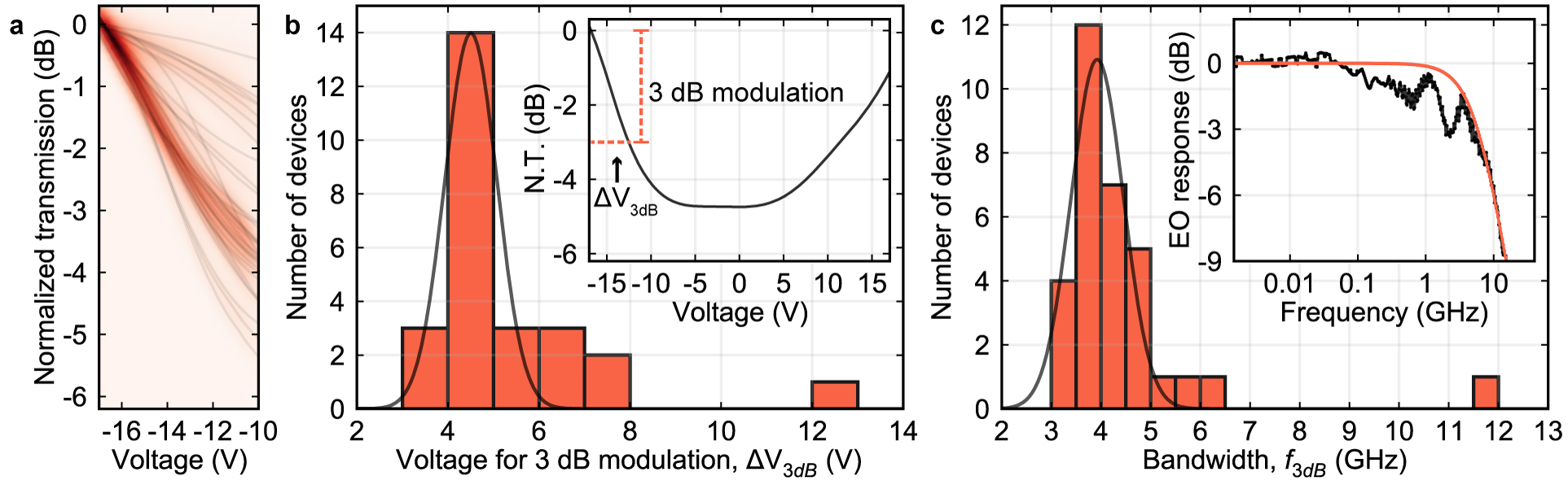}
        \caption{Statistics of graphene modulator performances.\\
        (a) Normalized transmission of 32 modulators as a function of voltage around the operating region (V = --17 V to --10 V). We superimpose the transmission density plot of all modulators (red) over the individual curve of each modulator (gray solid lines). The modulators exhibit similar transmission characteristics, especially around --17 V to --13 V, as shown by the dark and narrow width of the density curve. (b) Histogram of the measured voltage changes required for 3 dB modulation from maximum transmission, \vthreedb. Inset: example transmission curve describing \vthreedb. The solid line in the histogram is a normal distribution fit to the data with a standard deviation of 0.57 V (~$ < $~13\% of the mean 4.50 V). (c) Histogram of the measured modulator bandwidths. The standard deviation is 0.53 GHz (~$ < $~14\% of the mean 3.92 GHz). Inset: example frequency response curve along with the fit to a single-pole transfer function $ 1/(1 + j2\pi f \tau) $ (solid orange line), where $ f $ is the frequency and $ \tau $ is the modulator time constant.}
        \label{fig:dc}
    \end{figure}

	We show standard deviation of less than 13\% for both modulation efficiency (4.5 V for 3 dB modulation) and bandwidth (3.9 GHz) despite the devices exhibiting graphene residual doping variation of more than 30\% across the same chip area (8 mm $\times$ 1 mm, see \autoref{sfig:dirac} in the Supplementary Information). We show in \autoref{fig:dc}a the normalized transmission of 32 modulators with respect to voltage around the operating region (V = --17 V to --10 V). We superimpose the transmission density plot of all modulators (red) over the individual curve of each modulator (gray solid lines). The modulators exhibit similar transmission characteristics, especially around --17 V to --13 V, as shown by the dark and narrow width of the density curve. To quantify modulation efficiency, we measure the voltage required for 3 dB modulation from maximum transmission, \vthreedb, and plot a histogram in \autoref{fig:dc}b (the inset of \autoref{fig:dc}b shows an example transmission curve describing \vthreedb). We fit a normal distribution to the histogram (black solid line in \autoref{fig:dc}b) and measure a standard deviation of 0.57 V, which is less than 13\% of the mean (4.50 V). We also measure the modulators' bandwidths and plot the histogram in \autoref{fig:dc}c. We measure a standard deviation of 0.5 GHz from the fit, which is about 13\% from the mean (3.9 GHz). In the inset of \autoref{fig:dc}c, we show an example of the frequency response curve of one of the modulators. The orange line corresponds to the fit to a single-pole transfer function $ 1/(1 + j2\pi f \tau) $, where $ f $ is the frequency and $ \tau $ is the modulator time constant.

\begin{figure}[!t]
    	\centering
    	\noindent\makebox[\textwidth]{\includegraphics[width=180mm]{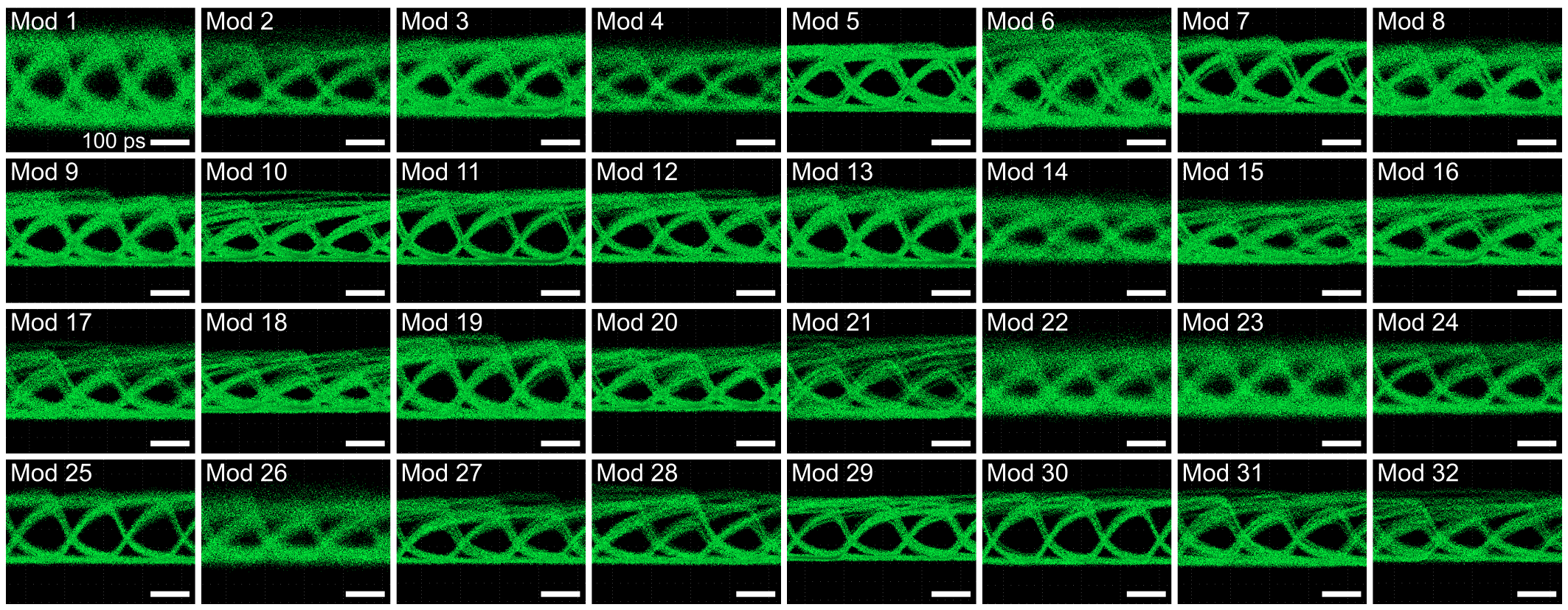}}
    	\caption{Measured eye diagrams of the graphene modulators at 7 \gbps.\\
    		Eye diagrams of each modulator driven with 2$ ^9 -1 $ \gls{nrz} \gls{prbs} at 7 \gbps\ and \vpp\ = 6 V. We achieve 94\% yield (30 out of 32 devices) with \gls{ber} below the \gls{hdfec} limit (\gls{ber} $<$ \SI{3.8e-3}{}) at 7 \gbps, amounting to a total data rate of 210 \gbps\ for the chip. The eye diagrams display similar eye opening and \gls{osnr} as implied by the narrow spread of the histogram of Q-factor in \autoref{sfig:eyeQ} of the Supplementary Information. The modulator number corresponds to that shown in the tiled optical micrograph in \autoref{fig:fabrication}b. The scale bars in all eye diagrams are 100 ps.}
    	\label{fig:eye}
    \end{figure}

    We measured 94\% yield (30 out of 32 devices) with \gls{ber} below the \gls{hdfec} limit (\gls{ber} $<$ \SI{3.8e-3}{}) \cite{agrawal2012fiber} at 7 \gbps, amounting to a total aggregated data rate of 210 \gbps\ for the chip. In \autoref{fig:eye} we show the measured eye diagram of each modulator driven with 2$ ^9 -1 $ \gls{nrz} \gls{prbs} at 7 \gbps\ and \vpp\ = 6 V (the modulator number corresponds to that shown in the tiled optical micrograph in \autoref{fig:fabrication}b).
	The eye diagrams display similar eye opening and \gls{osnr} as implied by the narrow spread of the histogram of Q-factor in \autoref{sfig:eyeQ} of the Supplementary Information. The small variation of both modulation efficiency (\vthreedb) and bandwidth ($ f_{3dB} $) enable modulators to transmit data with consistent performance across the chip at high speeds.

	The modulators currently consume about 1.6 pJ/bit (from $ E = CV_{\text{pp}}^2 / 4 $), which is mostly dissipated by the termination resistor. The modulator impedance is dominated by the small device capacitance of around 180 fF. To reduce reflections caused by the impedance mismatch between the capacitive load and 50 $ \Omega $ transmission line, we terminated the modulator using a second set of probes with a d.c. block capacitor and 50 $ \Omega $ RF termination. This shunt termination reduces the voltage drop on the modulator by approximately 50\%, so the applied voltage to the modulator is \vpp/2.

	\section{Design optimization towards Tbits/s data transmission}
\begin{figure}[!t]
		\centering
		\noindent\makebox[\textwidth]{\includegraphics[width=120mm]{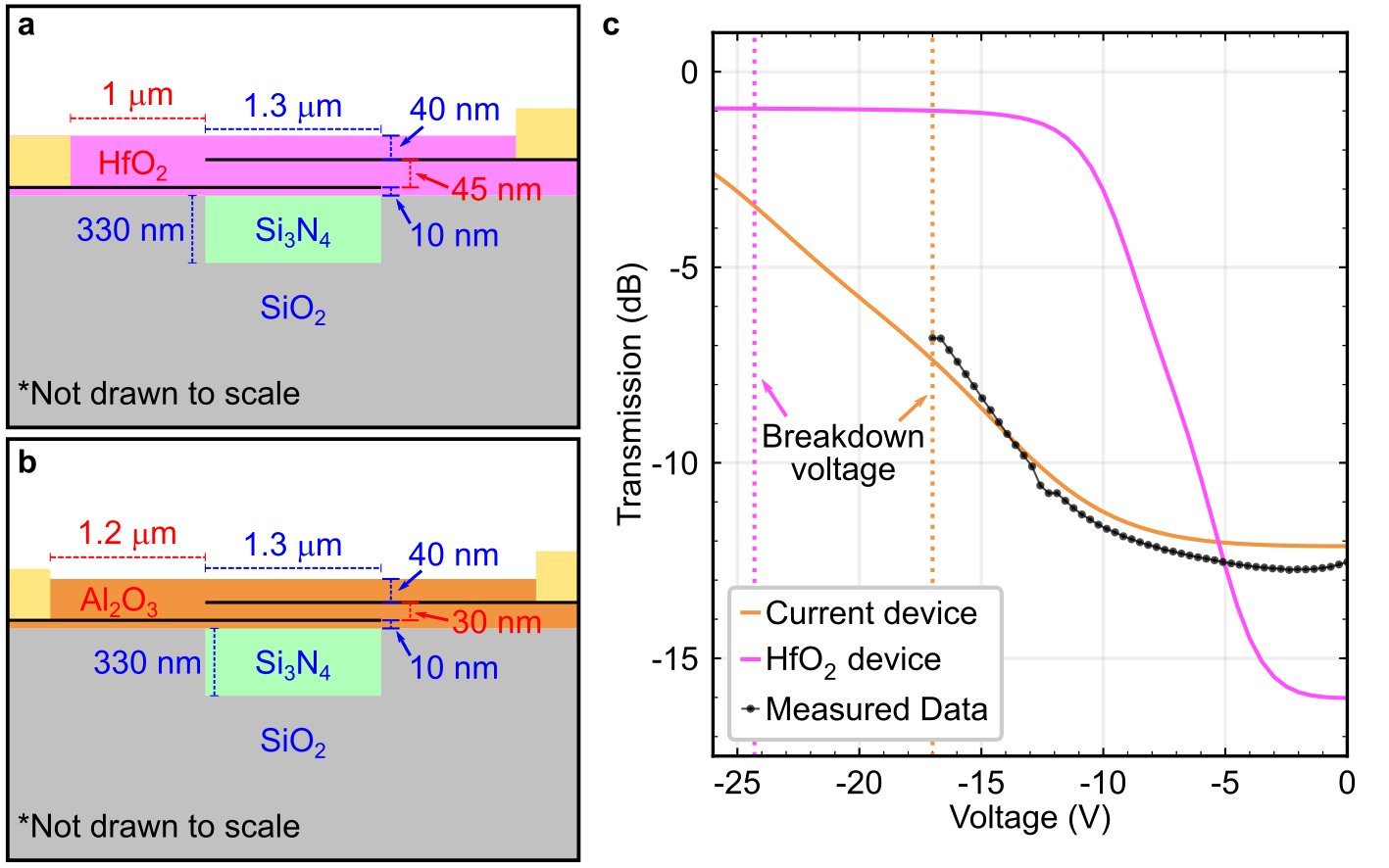}}
		\caption{Schematic of optimized device with expected transmission.\\
			(a) Optimized device cross-section with high-\kap\ gate. Here we simulate with \gls{hafnia}. The blue and red labels indicate unchanged and changed parameters between the two designs, respectively. (b) Current device cross-section. The \gls{sin} waveguide is 1,300 nm $ \times $ 330 nm and the graphene capacitor width is identical to the waveguide, 1.3 \um. The bottom encapsulating \gls{al2o3} is 10 nm, the gate \gls{al2o3} is 30 nm, and the top encapsulating \gls{al2o3} is 30 nm. The electrodes are 1.2 \um\ away from the waveguides. (c) The expected transmission curve versus applied voltage for the optimized design in (a) as purple curve and current devices in (b) as orange curve which is in close agreement with the measured data (black dots). One can see that with the optimized design (purple curve) we expect to achieve a modulation depth of 15 dB before breakdown with 1 dB insertion loss as it reaches further into Pauli-blocking, in contrast to our current device (orange curve) where we achieve a modulation of 5 dB with 7 dB insertion loss.}
		\label{fig:dev_sim}
	\end{figure}

	We show that with further optimization in device design, we can achieve stronger modulation extinction ratio and exceed the current aggregate data rate of 210 \gbps\ of the graphene transmitter chip. \autoref{fig:dev_sim}a shows a schematic of an optimized device with higher dielectric constant (\kap) based on \gls{hafnia} and, for comparison, \autoref{fig:dev_sim}b shows a schematic of our current modulator.
	The new design exhibits a stronger modulation extinction ratio due to both higher extrinsic doping and stronger overlap of the mode with the graphene. The extrinsic doping, \extdoping, is about \SI{1.4e13}{\per\centi\meter\squared} for \vpp\ = 6 V due to higher \kap\ dielectric, corresponding to six times the mean intrinsic doping while for our current device in \autoref{fig:dev_sim}b, \extdoping\ is about \SI{5.2e12}{\per\centi\meter\squared} for the same \vpp, corresponding to twice the intrinsic doping value. This stronger extrinsic doping in the optimized design leads to a greater contrast between graphene's absorptive and transparent state. In addition, the mode overlap with the graphene in the device shown in \autoref{fig:dev_sim}a is slightly higher, because \gls{hafnia} has higher refractive index than \gls{al2o3} (2.07 versus 1.63 at 1,550 nm, respectively), further contributing to stronger extinction ratio. In order to ensure that the larger \kap\ of \gls{hafnia} does not decrease the RC bandwidth, we design the device with an increased gate thickness from 30 nm to 45 nm and place the electrodes closer to the waveguide (1.0 \um\ compared to 1.2 \um\ for our current devices) to reduce graphene sheet resistance while adding negligible metal absorption. In \autoref{fig:dev_sim}c we show the expected transmission curve versus applied voltage for the new design (purple curve) and, for comparison, we also show the transmission curve for our current device (orange curve) which is in close agreement with the measured data (black dots). One can see that with the new design we expect to achieve a modulation depth of 15 dB before breakdown with 1 dB insertion loss as it reaches further into Pauli-blocking, in contrast to our current device where we achieve a modulation of 5 dB with 7 dB insertion loss.

\begin{figure}[!b]
		\centering
		\noindent\makebox[\textwidth]{\includegraphics[width=120mm]{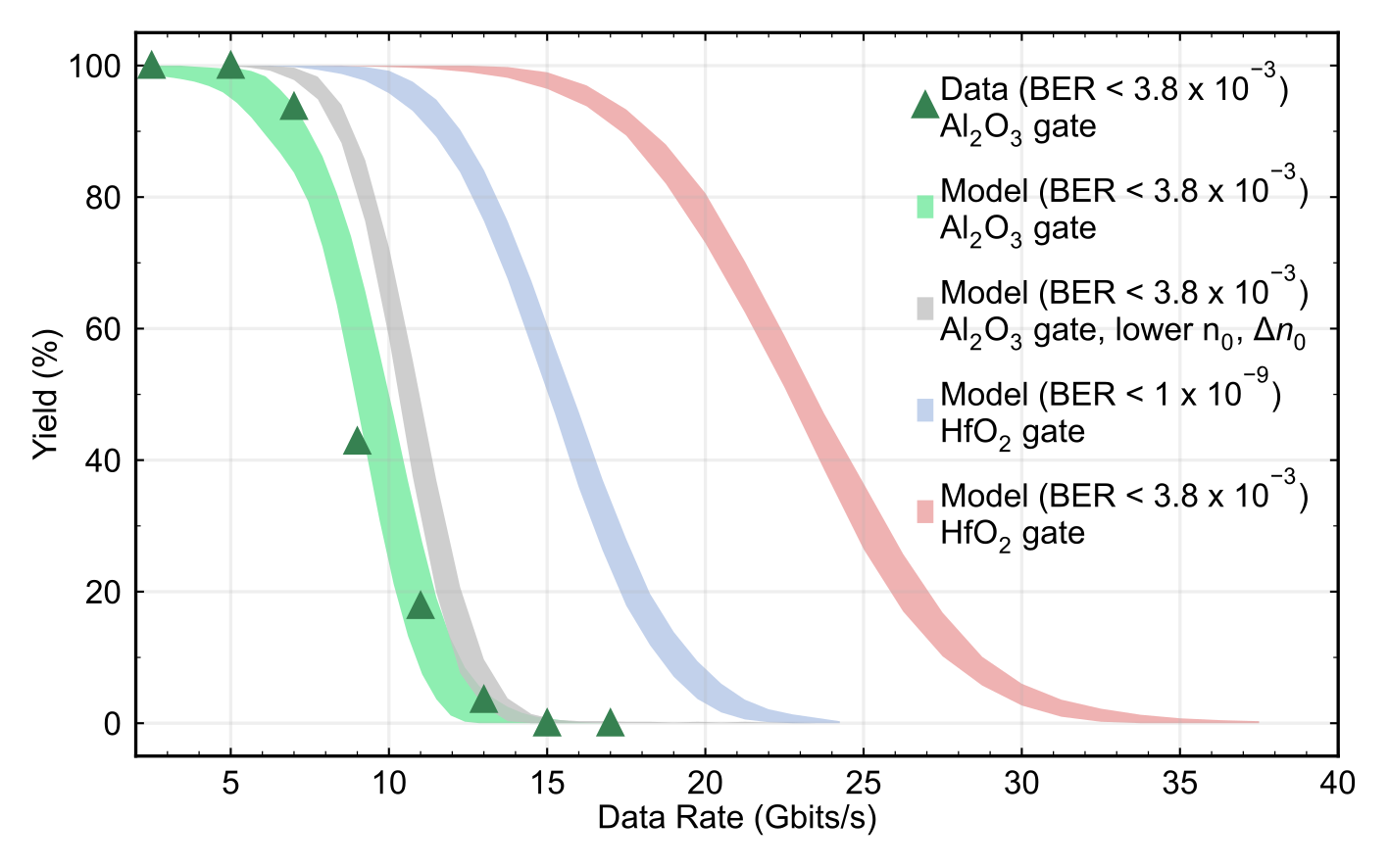}}
		\caption{Simulated yield versus data rate for devices in \autoref{fig:dev_sim} using Monte Carlo method\\
			Yield of devices in \autoref{fig:dev_sim}a and \autoref{fig:dev_sim}b as a function of data rate. The green triangle points correspond to the measured yield for \gls{ber}\ $ < $ \SI{3.8e-3}{}, and the colored curves are simulated yield curves where the upper and lower boundaries are 95\% interval (1.96 standard deviation). The green curve is the simulated yield for current device in (a) under variations of graphene doping measured in \autoref{sfig:dirac}, and is in good agreement with the measured data. The gray curve is the simulated yield for the current design with lower \doping\ = \SI{1e10}{\per\centi\meter\squared} and \deldoping\ = \SI{2.5e9}{\per\centi\meter\squared}. The red and blue curves are the simulated yields for the optimized high-\kap\ gate devices with \gls{hafnia} for \glspl{ber} $<$ \SI{3.8e-3}{} and \SI{1e-9}{}, respectively.}
		\label{fig:yield}
	\end{figure}

	We show with optimized design we can achieve $ > $ 80\% yield at data rate up to 20 \gbps, indicating scalability of the platform to higher data rates. We show that optimizing device cross-section has stronger effect on yield than lowering graphene doping, confirming that doping effects can be mitigated solely by device design. Using Monte Carlo method (see the Supplementary Information for the description of our model), we simulate yield as a function of data rate for the different device design as shown in \autoref{fig:yield}.
	The upper and lower boundaries of the curves are 95\% interval (1.96 standard deviation). We first confirm that the Monte Carlo model of our current device design with doping distribution from \autoref{sfig:dirac} (green curve in \autoref{fig:yield}) is in good agreement with the measured yield (green triangles) for \gls{ber} $ < $ \SI{3.8e-3}{} (also see \autoref{sfig:ber} in the Supplementary Information for \gls{ber} $ < $ \SI{1e-4}{}). The simulated yield for the optimized high-\kap\ gate devices and similar distribution of graphene doping as current modulators for \gls{ber} $ < $ \SI{3.8e-3}{} and \SI{1e-9}{} (error-free for certain applications \cite{stern2015chip, agrawal2012fiber}) are shown in the red and blue curves, respectively. In the gray curve of \autoref{fig:yield}, we show the simulated yield for a device with \gls{al2o3} (same as current device) but with lower level of graphene doping and its variation, \doping\ = \SI{1e10}{\per\centi\meter\per\squared} and \deldoping\ = \SI{2.5e9}{\per\centi\meter\squared}, respectively, which are reported values for graphene encapsulated with BN \cite{dean2010boron, xue2011scanning}. One can see that despite having larger distribution of graphene doping, the optimized devices with high-\kap\ exhibit yield close to 100\% at 12.5 \gbps\ at \gls{ber} $ < $ \SI{3.8e-3}{} (red curve) in contrast to 12\% yield at same data rate and \gls{ber} of our current devices (gray curve). It is also evident that the optimized high-\kap\ devices exhibit greater yield (close to 85\%) even with stricter \gls{ber} $ < $ \SI{1e-9}{} (blue curve) than current devices with assumed lower doping (gray curve), suggesting larger extrinsic doping improves data transmission more significantly than lowering \doping\ and \deldoping.

    \section{Conclusion}
    We have demonstrated that optimizing graphene-mode overlap and device capacitance of graphene-based modulators can minimize the effect of large inhomogeneities of \gls{cvd} graphene films on device yield. We have achieved small variation in modulation efficiency and bandwidth for our modulators that allow them to transmit data with high fidelity. We also demonstrate the feasibility of data rates approaching Tbits/s with graphene modulators by further optimizing mode confinement and device capacitance using high-\kap\ gate dielectric such as \gls{hafnia}. This demonstration places graphene as the electro-optic material capable of simultaneously supporting high bandwidth, consuming low power, and achieving high integration density. In addition, as these modulators are not based on resonant cavities, they are capable of broadband and athermal operation \cite{dalir2016athermal}, which paves the way for graphene-based wavelength and spatial division multiplexing to further enhance data transmission capacity.

    \section*{Acknowledgment}
    The authors would like to thank Dr. Aseema Mohanty for the fruitful discussions and experimental support. This work was performed in part at the City University of New York Advanced Science Research Center NanoFabrication Facility and in part at the Columbia Nano Initiative (CNI) shared labs at Columbia University in the City of New York.

    \section*{Funding}
    We gratefully acknowledge support Programmable Quantum Materials, an Energy Frontier Research Center funded by the U.S. Department of Energy (DOE), Office of Science, Basic Energy Sciences (BES), under award DE-SC0019443.

    % ---- Bibliography
    \bibliographystyle{unsrt}
    \bibliography{References}

    \clearpage
    \setcounter{page}{1}
    \setcounter{figure}{0}
    \setcounter{section}{0}
    \setcounter{equation}{0}

    \renewcommand{\thepage}{S\arabic{page}}
    \renewcommand{\thefigure}{S\arabic{figure}}
    \renewcommand{\thetable}{S\arabic{table}}
    \def\theequation{S\arabic{equation}}

    % ---------- Supplementary information ----------

    \begin{center}
        \Huge \textbf{Supplementary Information}
    \end{center}

	\section{Graphene transfer via electrochemical delamination}
\begin{figure}[!ht]
        \centering
        \includegraphics[width=143mm]{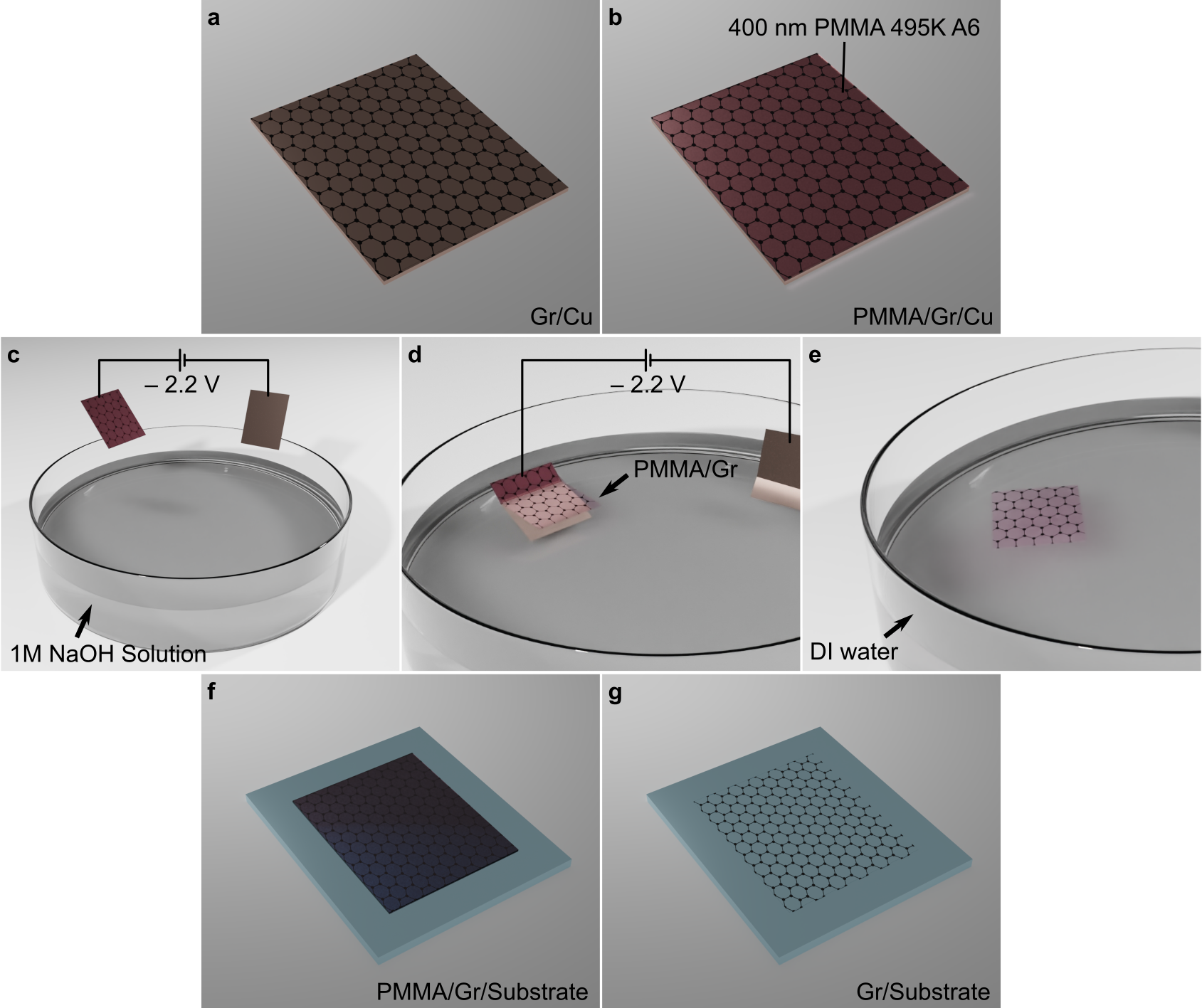}
        \caption{Schematics of graphene transfer via electrochemical delamination.\\
        	(a) We start with large area (75 mm $ \times $ 75 mm) \gls{cvd} graphene on copper substrate (25 \um-thick) grown by Grolltex Inc. \protect\citeSup{supgrolltex}. (b) We spin coat 400 nm-thick PMMA 495K A6 on graphene to provide mechanical support during the delamination. We dry the PMMA layer overnight at ambient conditions without additional baking steps. (c) For the electrolyte we prepare 1M \ch{NaOH} solution. The PMMA/graphene/Cu foil acts as a cathode and another bare Cu foil as an anode. (d) We apply --2.2 V to the graphene sheet with respect to the copper anode, and slowly submerge both the anode and PMMA/graphene/Cu cathode into the electrolyte. The PMMA/graphene stack begins to delaminate due to ion intercalation effect \protect\citeSup{supverguts2018graphene} and floats due to surface tension. The delamination takes about 10 - 15 seconds for film size around 20 mm x 20 mm. (e) We transfer the floating PMMA/graphene stack to a fresh de-ionized (DI) water bath using a glass slide to rinse the electrolyte. We rinse the PMMA/graphene stack in two DI water baths, 5 minutes in each bath. (f) We transfer the graphene film onto a substrate with flat surface pre-treated with \ch{O2} plasma. We dry the wet substrate in a vacuum desiccator with a base pressure of around 0.5 Torr for at least 24 hours to pump out residual water. (g) We bake the sample on a hot plate at 180 \si{\celsius} for two hours and strip the PMMA film in acetone bath for at least 1 hour.}
        \label{sfig:transfer}
    \end{figure}

	We describe the electrochemical delamination transfer in \autoref{sfig:transfer}. We start with large area (75 mm $ \times $ 75 mm) \gls{cvd} graphene on copper substrate (25 \um-thick) grown by Grolltex Inc. \citeSup{supgrolltex}. We spin coat 400 nm-thick PMMA 495K A6 on graphene to provide mechanical support during the delamination. We dry the PMMA layer overnight at ambient conditions without additional baking steps. For the electrolyte we prepare 1M \ch{NaOH} solution. The PMMA/graphene/Cu foil acts as a cathode and another bare Cu foil as an anode. We apply --2.2 V to the graphene sheet with respect to the copper anode, and slowly submerge both the anode and PMMA/graphene/Cu cathode into the electrolyte. The PMMA/graphene stack begins to delaminate due to ion intercalation effect \citeSup{supverguts2018graphene} and floats due to surface tension. The delamination takes about 10 - 15 seconds for film size around 20 mm x 20 mm. We then transfer the floating PMMA/graphene stack to a fresh de-ionized (DI) water bath using a glass slide to rinse the electrolyte. We rinse the PMMA/graphene stack in two DI water baths, 5 minutes in each bath. We transfer the graphene film onto a substrate with flat surface pre-treated with \ch{O2} plasma. Cleaning the substrate with \ch{O2} plasma strips residues and makes the surface hydrophilic, which facilitates the removal of trapped water between graphene and the surface during the drying step. We dry the wet substrate in a vacuum desiccator with a base pressure of around 0.5 Torr for at least 24 hours to pump out residual water. Fully drying the sample significantly reduced tears in the graphene. After vacuum-drying the sample, we bake it on a hot plate at 180 \si{\celsius} for two hours and strip the PMMA film in acetone bath for at least 1 hour.

    \section{Modeling device yield with Monte Carlo Simulations}
    \subsection{Screening influential parameters with sensitivity analysis}
	We first perform sensitivity analysis with \gls{eet} method \citeSup{morris1991factorial} to identify device parameters that have strong influence on extinction ratio. By identifying the most influential parameters, we focus on varying these variables while fixing others constant in our Monte Carlo simulations to save computational cost \citeSup{waqas2017sensitivity}.

\begin{figure}[!h]
		\centering
		\includegraphics[width=100mm]{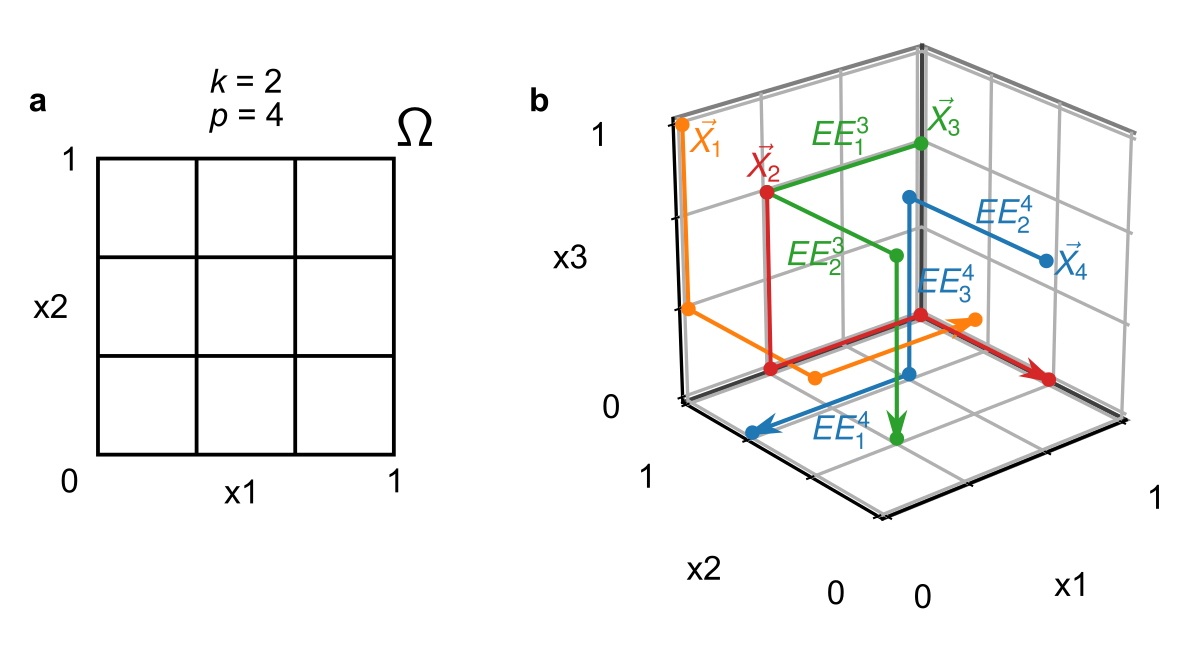}
		\caption{Illustration of discretizing the input space. (a) Input space $\Omega$ with $ k=2 $ and $ p = 4 $. Each variable, x1 and x2, are divided into $ p=4 $ levels. (b) Input space $\Omega$ with $ k=3 $ and $ p=4 $. We sample $ r=4 $ input vectors or initial points, $ \vec{X_1} $ to $ \vec{X_4} $. Each point progresses through a trajectory in a 'one-at-a-time' fashion, where each variable randomly and sequentially changes by an increment of $\Delta$. Through each movement, we compute \gls{ee}. We perform sensitivity analysis by computing mean and standard deviations of these \glspl{ee}.}
		\label{sfig:kp_grid}
	\end{figure}

	\gls{eet} method is based on calculating a number of differential ratios, called \glspl{ee}, for a given input vector, from which statistics are computed to derive sensitivity information. To illustrate \gls{eet}, let us assume there are $ k $ input variables, such as waveguide width, thickness, etc., and $ \vec{X} = (X_1, X_2, ..., X_k) $ is the input vector. Next, we assume that each parameter is varied across $ p $-levels, such that the whole $ k $-dimensional input space is discretized as a $ p^k $-point grid, which we call it region $ \Omega $. We show $\Omega$ with $ k=2 $ and $ p = 4 $ in \autoref{sfig:kp_grid}a, and $\Omega$ with $ k=3 $ and $ p = 4 $ in \autoref{sfig:kp_grid}b as examples. Let $ Y $ be the output of interest (e.g. extinction ratio, insertion loss, etc.) from a model $ f(\vec{X}) $,
	\begin{equation}
		Y = f(X_1, X_2, ..., X_k).
	\end{equation}
	The gist of \gls{eet} is to randomly sample $ r \in \{1, 2, ...\}$ number of input vectors or initial points ($\vec{X_1}$, $\vec{X_2}$, ..., $\vec{X_r} \in  \Omega $) and then move these points in $ \Omega $ by randomly changing each variable in one-at-a-time fashion, creating a ($ k+1 $)-point trajectory for each initial point. We illustrate this with $ r=4 $ input points ($\vec{X_1}$, ..., $\vec{X_4}$) sampled in $ \Omega $ with $ k=3 $ and $ p=4 $ grid in \autoref{sfig:kp_grid}b. Each input point randomly moves in a basis direction ($ \hat{x}_1 $, $ \hat{x}_2 $, or $ \hat{x}_3 $ in \autoref{sfig:kp_grid}b), and during each step through the trajectory we calculate an \gls{ee} defined by:
	\begin{equation}
		EE^j_i(\vec{X}) = \frac{Y(X_1, ..., X_i + \Delta, ..., X_k) - Y(\vec{X})}{\Delta}
	\end{equation}
	where $ j \in {1, ..., r} $, $ i $ is the variable being moved ($ i \in {1, ..., k} $), and $\Delta$ is incremental step chosen from the set $ \{0, 1/(p-1), 2/(p-1), ... , 1\} $ (in \autoref{sfig:kp_grid}b, for example, we choose $\Delta = 2/(p-1) = 2/3$). Some \glspl{ee} are labeled in \autoref{sfig:kp_grid}b to their corresponding edges. For a given $ \Omega $, sufficient number of samples, $ r $, should be chosen such that the trajectories effectively cover $ \Omega $ without trading off computational cost.
	Once all the \glspl{ee} are computed for all trajectories taken by randomly sampled input vectors, for each input factor $ i $ we measure $ \mu_i $ and $ \sigma_i $, which are mean and standard deviation of \glspl{ee} related to input $ i $, defined as \citeSup{campolongo2007effective}:
	\begin{align}
		&\mu_i = \frac{1}{r}\sum_{j=1}^{r} | EE_i^j | = \frac{1}{r}\left( |EE_i^1| + |EE_i^2| + ... + |EE_i^r|\right) \\
		&\sigma_i^2 = \frac{1}{r-1} \sum_{j=1}^{r} (EE_i^j - \mu_i)^2.
	\end{align}
	These statistical parameters of \glspl{ee} provide a direct qualitative knowledge on the influence of input $ i $ on the model. The mean, $\mu_i$ assesses the overall influence of a parameter $ i $ on the model output, and $\sigma_i$ describes non-linear effects \citeSup{morris1991factorial}. A large $ \sigma_i $ indicates that the parameter $ i $ is interacting with others in a non-linear way because its \glspl{ee} vary greatly depending on which trajectories they originate from.

\begin{figure}[!h]
		\centering
		\includegraphics[width=52mm]{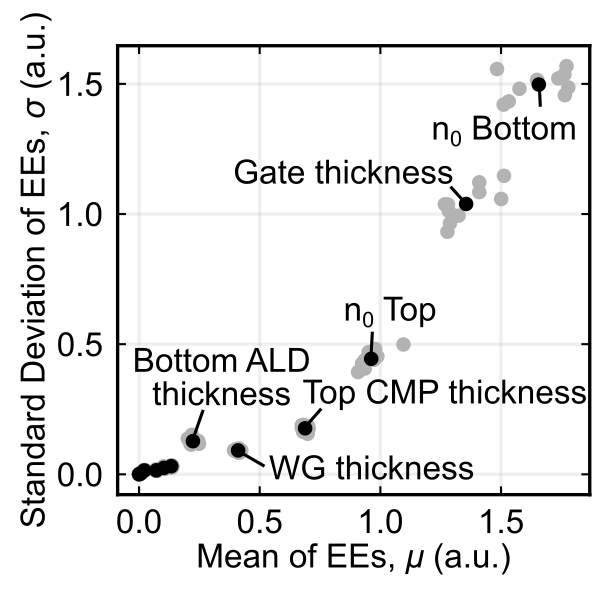}
		\caption{Calculation of $\mu$ and $\sigma$ of \glspl{ee} for our modulator design with $ k=12 $ variables, $ p = 4 $ levels, $ r = 100 $ sampling points for [xx] iterations. Black dots are average values of $ \mu $ and $\sigma$ of each parameter from [45] iterations. In the background we superimpose all the points from [45] iterations as gray dots to show the full statistics. The unlabeled parameters near the origin are: graphene capacitor width, top ALD cladding thickness, waveguide width, dielectric constant, waveguide side-wall angle, and metal-waveguide gap, in decreasing magnitude of $\mu$.}
		\label{sfig:morris}
	\end{figure}

	We perform \gls{eet} for our modulator design, which has a total of 12 parameters, and determine that graphene residual doping (for both top and bottom sheets) and gate thickness are the most influential parameters on extinction ratio as they have the greatest $\mu$. As expected, parameters that directly relate to Pauli-blocking in graphene and control the degree of graphene-mode overlap are the most influential parameters. Moreover, since bottom graphene sheet has stronger interaction with the mode, its level of doping is expected to have stronger influence on extinction ratio than the top graphene doping. For subsequent Monte Carlo iterations, we fix other parameters constant at their designed values to save computational cost. We run [45] independent studies and plot the mean and standard deviations of \gls{ee} for each parameter in \autoref{sfig:morris}. The black dots are average values of $ \mu $ and $\sigma$ from [45] iterations. In the background we superimpose all the points from [45] iterations as gray dots to show the full statistics.

	\subsection{Implementation of Monte Carlo simulations}
\begin{figure}[!h]
		\centering
		\includegraphics[width=66.5mm]{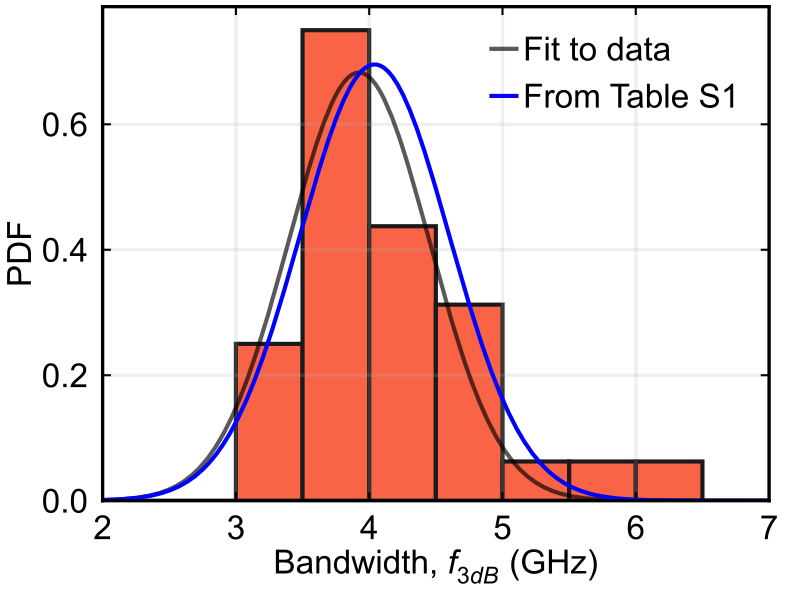}
		\caption{Comparison between calculated bandwidth based on measured variation of RC parameters and the data. The histogram is measured bandwidth from \autoref{fig:dc}c. Black solid line is the normal distribution fit to the measured data. Blue solid line is bandwidth distribution calculated based on measured RC parameters in \autoref{stab:variation} and $ R_{sh} $ (from \doping\ and $ \mu_{Hall} $) and $ R_c $. The calculated mean and standard deviation is 4.0 GHz and 0.56 GHz, respectively. The calculated distribution of bandwidth is in good agreement with the measurement.}
		\label{sfig:bandwidth}
	\end{figure}

\begin{figure}[!ht]
        \centering
        \includegraphics[width=60mm]{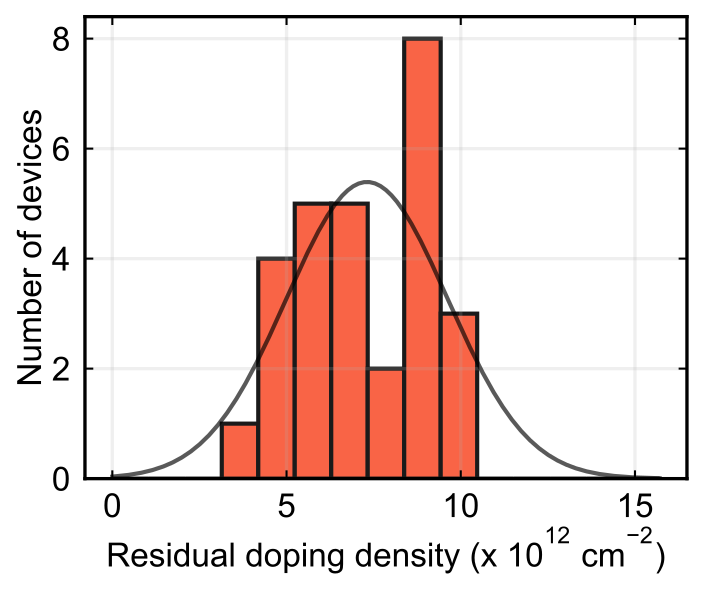}
        \caption{Histogram of graphene residual doping measured using \glspl{gfet}.\\
        We characterize residual doping in graphene sheets prepared with electrochemical delamination (see \autoref{sfig:transfer}). We measure the Dirac point of graphene channels in \glspl{gfet} fabricated evenly throughout a test chip of similar size to our graphene transmitter chip (8 mm $\times$ 1 mm). We sweep the gate voltage and measure the voltage that yields minimum drain-source current. We calculate the residual doping from this voltage, $ V_{CNP} $ by $ n = CV_{CNP}/e $, where C is the measured \gls{gfet} capacitance of \SI{56}{\nano\farad\per\centi\meter\squared} and $ e $ is the positive elementary charge. The solid line is the normal distribution fit to the data with mean \doping\ = \SI{7.3e12}{\per\centi\meter\squared} and standard deviation \deldoping\ = \SI{2.3e12}{\per\centi\meter\squared} (32\% of mean).}
        \label{sfig:dirac}
    \end{figure}

	We perform Monte Carlo simulations based on statistical measurements of the most influential parameters on extinction ratio identified using \gls{eet} sensitivity analysis. These are graphene residual doping and gate dielectric thickness, and measurements of these parameters are summarized in \autoref{stab:variation}. We measure graphene residual doping, \doping, by measuring charge neutral points, $ V_{CNP} $, of 28 \glspl{gfet} evenly placed over similar chip area to our transmitter chip (8 mm $ \times $ 1 mm), and calculate \doping\ via the relation $ n_0 = CV_{CNP}/e $, where the capacitance is measured via graphene Hall bar structures. We plot residual doping histogram in \autoref{sfig:dirac}, whose measured mean and standard deviation from normal distribution fit are \doping\ = \SI{7.3e12}{\per\centi\meter\squared} and \deldoping\ = \SI{2.3e12}{\per\centi\meter\squared}, respectively. For the gate dielectric thickness we measure a standard deviation of 2 nm. We measure the thickness variation by depositing 30 nm (target thickness) \gls{ald} \gls{al2o3} on silicon substrate and measure the thickness of film at various points (across 8 mm $ \times $ 1 mm area) with an ellipsometer. Dielectric constant is deduced from capacitance measurements using a graphene Hall bar as \kap\ = 4.2.
	\begin{table}[!h]
		\centering
		\begin{tabular}{|c|c|c|}
			\hline
			Parameter & Mean & Std \\
			\hline
			n0 ($\times 10^{12}$ cm$^{-2}$) & 7.3 & 2.3 \\
			\hline
			Gate thickness (nm) & 29.1 & 2.0 \\
			\hline
		\end{tabular}
		\caption{Measured variation of parameters that significantly affect modulator extinction ratio. We measure graphene residual doping, \doping, by measuring charge neutral points, $ V_{CNP} $, of 28 \glspl{gfet} evenly placed over similar chip area to our transmitter chip (8 mm $ \times $ 1 mm), and calculate \doping\ via the relation $ n_0 = CV_{CNP}/e $. To measure variations in gate dielectric thickness we deposit 30 nm (target thickness) \gls{ald} \gls{al2o3} on silicon substrate and measure the thickness of film at various points (across 8 mm $ \times $ 1 mm area) with an ellipsometer.}
		\label{stab:variation}
	\end{table}
	We also measure contact resistance and Hall mobility (for sheet resistance) of $ R_c $ = 10.5 $ \pm $ 1.3 k\ohmum\ and $ \mu $ = 1420 $\pm$ 26 \cmvs , respectively. We finally note that our measured variations of RC related parameters (\autoref{stab:variation}, $ R_c $, $\mu_{Hall}$, etc.) provide similar distribution of calculated bandwidth as our measured data in \autoref{fig:dc}c, as shown in \autoref{sfig:bandwidth}. The bandwidth calculated from variations of RC related parameters (blue solid curve) has similar mean and standard deviation (mean = 4.0 GHz, std = 0.56 GHz) as the data and its fit (black solid curve).

\begin{figure}[!t]
		\centering
		\noindent\makebox[\textwidth]{\includegraphics[width=183mm]{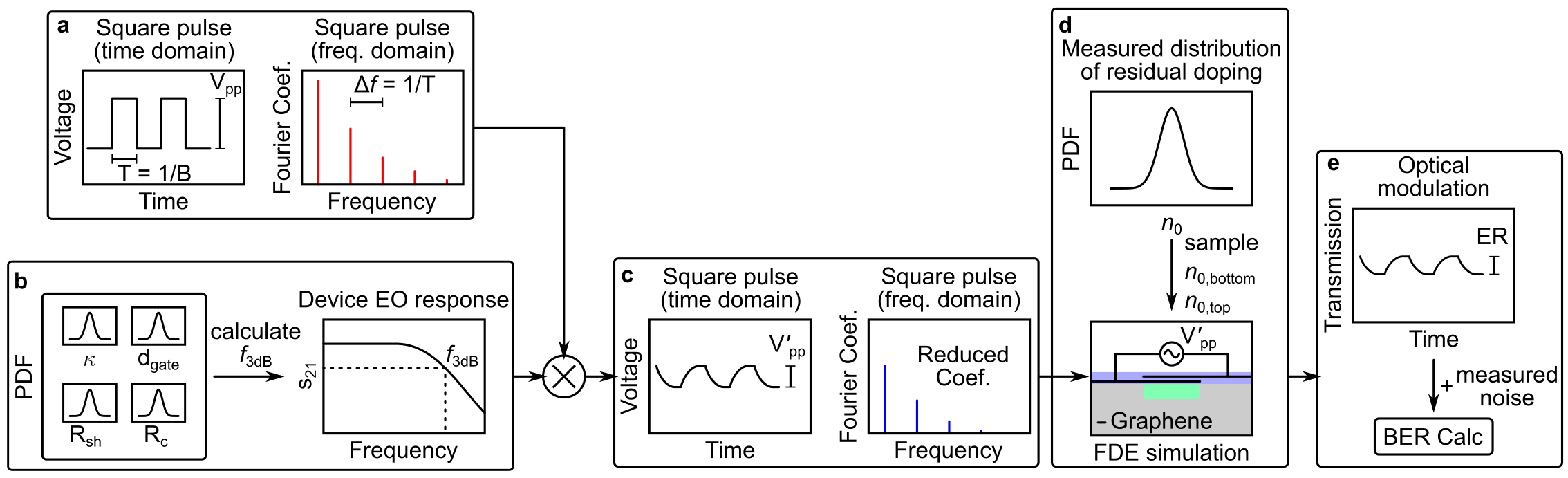}}
		\caption{Monte Carlo simulation.\\
			(a) We construct the input voltage square pulses (to emulate \gls{nrz} \gls{prbs}) with peak-to-peak amplitude equal to the drive voltage reduced by 50 $ \Omega $ termination (\vpp/2) in the time domain at a given data rate, $ B $, and find its Fourier coefficients. (b) We calculate the device RC bandwidth by sampling RC related parameters -- dielectric constant (\kap), gate thickness ($ d_{gate} $), sheet resistance ($ R_{sh} $), contact resistance ($ R_c $) -- and calculate $ f_{3dB} $ and resulting $ S_{21} $ as a function of frequency. (c) We multiply the Fourier coefficients of the square pulse with the frequency response of the device to calculate the new pulse applied to the device. This new amplitude ($ V'_{pp} $) is applied to the modulator. (d) We sample graphene residual doping for both sheets of graphene in the capacitor and gate thickness from (a) to create a new device geometry, and calculate the extinction ratio with the drive voltage ($ V'_{pp} $). with a \gls{fde} solver. (e) We use the calculated extinction ratio from (d) in order to calculate the quality factor and \gls{ber} according to \autoref{seq:quality}. }
		\label{sfig:monte}
	\end{figure}

	We describe the flow of our Monte Carlo simulations for emulating device performance and data transmission yield in \autoref{sfig:monte}. We first construct the input voltage square pulses in the time domain at a given data rate, $ B $, and find its Fourier coefficients (\autoref{sfig:monte}a). We subsequently calculate device bandwidth by sampling RC related parameters from our measured distributions and compute $ s_{21} $ as a function of frequency (\autoref{sfig:monte}b). We multiply $ s_{21} $ with Fourier coefficients of the input signal to deduce the voltage applied to the modulator,$ V'_{pp} $, limited by its bandwidth (\autoref{sfig:monte}c). We construct the modulator cross-section using sampled doping concentration for bottom and top graphene and gate thickness as shown in \autoref{sfig:monte}d. Then, we extract the extinction ratio due to $ V'_{pp} $ with a \gls{fde} solver (Lumerical \citeSup{lumerical}) and calculate the \gls{ber}.

	We compute \gls{ber} by first calculating Q-factor \citeSup{supagrawal2012fiber}:
	\begin{equation}\label{seq:quality}
		Q = \frac{P_1 - P_0}{\sigma_1 + \sigma_0}
	\end{equation}
	where $ (P_1, \sigma_1) $ and $ (P_0, \sigma_0) $ are optical power and RMS noise of level 1 and level 0, respectively, measured by the receiver, and $ r $ is the extinction ratio, $ r = P_1 / P_0 $, and $ \sigma^* = (\sigma_1 + \sigma_0)/P_0 $ is the total RMS noise normalized to level 0. We measure average $ \sigma^* $ of [0.11], which we use in the Monte Carlo simulations. Once we calculate the Q-factor, the \gls{ber} is equal to \gls{ber} = 0.5erfc(Q/$\sqrt{2}$) \cite{agrawal2012fiber}, where erfc is the complementary error function.
	
	\clearpage
    \section{Supplementary Figures}

    \begin{figure}[!ht]
        \centering
        \includegraphics[width=60mm]{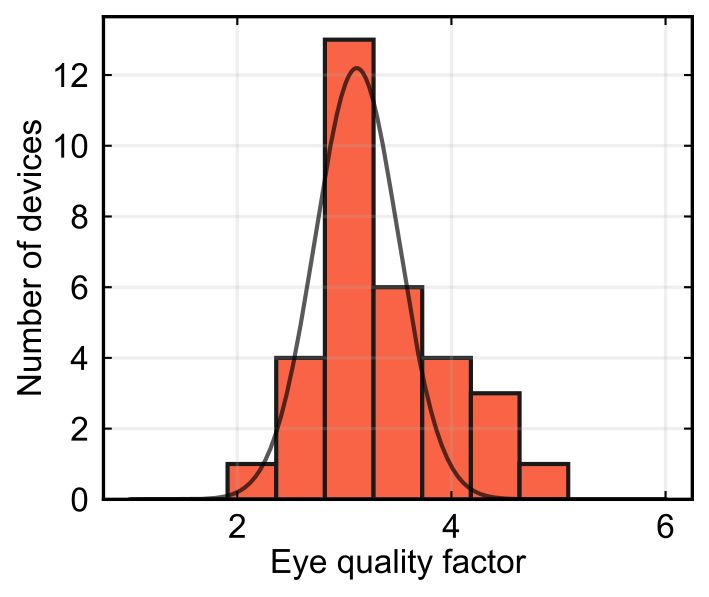}
        \caption{Histogram of eye diagram quality factors in \autoref{fig:eye} at 7 \gbps. Mean = 3.12, standard deviation = 0.39 (13\% of mean).}
        \label{sfig:eyeQ}
    \end{figure}

    \begin{figure}[!ht]
        \centering
        \includegraphics[width=100mm]{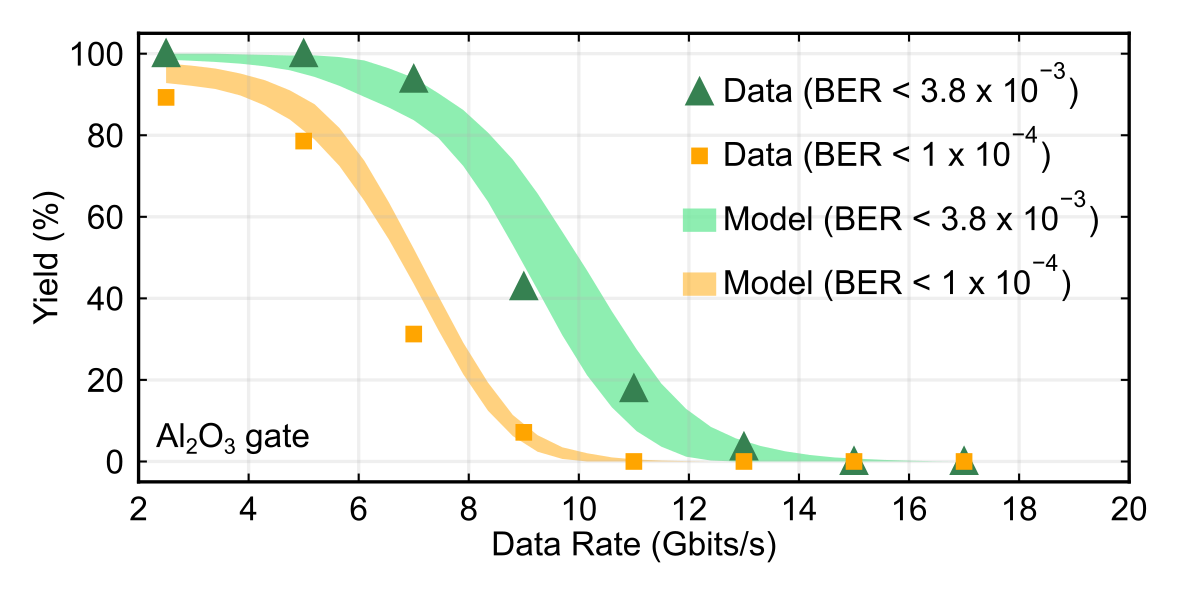}
        \caption{Measured and simulated device yield with respect to data rate for different \glspl{ber}. The Monte Carlo model described in the Supplementary Information emulates the yield curves at \glspl{ber} $ < $ \SI{3.8e-3}{} and \SI{1e-4}{}.}
        \label{sfig:ber}
    \end{figure}

    \clearpage
    \bibliographystyleSup{unsrt}
    \bibliographySup{SReferences}

\end{document}